\RequirePackage{lineno}
\documentclass[%
 reprint,
 superscriptaddress,
showpacs,preprintnumbers,
amsmath,
amssymb,
aps,
prl,
floatfix,
]{revtex4-1}

\usepackage{graphicx}% Include figure files
\usepackage{epstopdf}
\usepackage{dcolumn}% Align table columns on decimal point
\usepackage{bm}% bold math
\usepackage{color}
\usepackage{xspace}
\usepackage{hyperref}% add hypertext capabilities
\usepackage{natbib}
\usepackage{ulem}
\usepackage{multirow}
\input symbols
\linenumbersep{0.1cm}
\begin{document}

%\linenumbers

%\preprint{T2K DRAFT FINAL VERSION (v4)}

\title{Observation of Electron Neutrino Appearance in a Muon Neutrino Beam}

%%%%%%%%%%%%%%%%%%%%%%%%%%%%%%%%%%%%%%%%%%%%%%%%%%%%%%%%%%%%%%
% T2K author list generated on Tue, 12 Nov 2013 15:06:37 +0900
% setting: extra = 0 revtex = 1 yearrule = 1 shiftrule = 1
%         shift rules based on: period1 = 1112 period2 = 1213
% Number of authors = 338
%%%%%%%%%%%%%%%%%%%%%%%%%%%%%%%%%%%%%%%%%%%%%%%%%%%%%%%%%%%%%%

\newcommand{\INSTC}{\affiliation{University of Alberta, Centre for Particle Physics, Department of Physics, Edmonton, Alberta, Canada}}
\newcommand{\INSTEE}{\affiliation{University of Bern, Albert Einstein Center for Fundamental Physics, Laboratory for High Energy Physics (LHEP), Bern, Switzerland}}
\newcommand{\INSTFE}{\affiliation{Boston University, Department of Physics, Boston, Massachusetts, U.S.A.}}
\newcommand{\INSTD}{\affiliation{University of British Columbia, Department of Physics and Astronomy, Vancouver, British Columbia, Canada}}
\newcommand{\INSTGA}{\affiliation{University of California, Irvine, Department of Physics and Astronomy, Irvine, California, U.S.A.}}
\newcommand{\INSTI}{\affiliation{IRFU, CEA Saclay, Gif-sur-Yvette, France}}
\newcommand{\INSTGB}{\affiliation{University of Colorado at Boulder, Department of Physics, Boulder, Colorado, U.S.A.}}
\newcommand{\INSTFG}{\affiliation{Colorado State University, Department of Physics, Fort Collins, Colorado, U.S.A.}}
\newcommand{\INSTFH}{\affiliation{Duke University, Department of Physics, Durham, North Carolina, U.S.A.}}
\newcommand{\INSTBA}{\affiliation{Ecole Polytechnique, IN2P3-CNRS, Laboratoire Leprince-Ringuet, Palaiseau, France }}
\newcommand{\INSTEF}{\affiliation{ETH Zurich, Institute for Particle Physics, Zurich, Switzerland}}
\newcommand{\INSTEG}{\affiliation{University of Geneva, Section de Physique, DPNC, Geneva, Switzerland}}
\newcommand{\INSTDG}{\affiliation{H. Niewodniczanski Institute of Nuclear Physics PAN, Cracow, Poland}}
\newcommand{\INSTCB}{\affiliation{High Energy Accelerator Research Organization (KEK), Tsukuba, Ibaraki, Japan}}
\newcommand{\INSTED}{\affiliation{Institut de Fisica d'Altes Energies (IFAE), Bellaterra (Barcelona), Spain}}
\newcommand{\INSTEC}{\affiliation{IFIC (CSIC \& University of Valencia), Valencia, Spain}}
\newcommand{\INSTEI}{\affiliation{Imperial College London, Department of Physics, London, United Kingdom}}
\newcommand{\INSTGF}{\affiliation{INFN Sezione di Bari and Universit\`a e Politecnico di Bari, Dipartimento Interuniversitario di Fisica, Bari, Italy}}
\newcommand{\INSTBE}{\affiliation{INFN Sezione di Napoli and Universit\`a di Napoli, Dipartimento di Fisica, Napoli, Italy}}
\newcommand{\INSTBF}{\affiliation{INFN Sezione di Padova and Universit\`a di Padova, Dipartimento di Fisica, Padova, Italy}}
\newcommand{\INSTBD}{\affiliation{INFN Sezione di Roma and Universit\`a di Roma ``La Sapienza'', Roma, Italy}}
\newcommand{\INSTEB}{\affiliation{Institute for Nuclear Research of the Russian Academy of Sciences, Moscow, Russia}}
\newcommand{\INSTHA}{\affiliation{Kavli Institute for the Physics and Mathematics of the Universe (WPI), Todai Institutes for Advanced Study, University of Tokyo, Kashiwa, Chiba, Japan}}
\newcommand{\INSTCC}{\affiliation{Kobe University, Kobe, Japan}}
\newcommand{\INSTCD}{\affiliation{Kyoto University, Department of Physics, Kyoto, Japan}}
\newcommand{\INSTEJ}{\affiliation{Lancaster University, Physics Department, Lancaster, United Kingdom}}
\newcommand{\INSTFC}{\affiliation{University of Liverpool, Department of Physics, Liverpool, United Kingdom}}
\newcommand{\INSTFI}{\affiliation{Louisiana State University, Department of Physics and Astronomy, Baton Rouge, Louisiana, U.S.A.}}
\newcommand{\INSTJ}{\affiliation{Universit\'e de Lyon, Universit\'e Claude Bernard Lyon 1, IPN Lyon (IN2P3), Villeurbanne, France}}
\newcommand{\INSTCE}{\affiliation{Miyagi University of Education, Department of Physics, Sendai, Japan}}
\newcommand{\INSTDF}{\affiliation{National Centre for Nuclear Research, Warsaw, Poland}}
\newcommand{\INSTFJ}{\affiliation{State University of New York at Stony Brook, Department of Physics and Astronomy, Stony Brook, New York, U.S.A.}}
\newcommand{\INSTGJ}{\affiliation{Okayama University, Department of Physics, Okayama, Japan}}
\newcommand{\INSTCF}{\affiliation{Osaka City University, Department of Physics, Osaka, Japan}}
\newcommand{\INSTGG}{\affiliation{Oxford University, Department of Physics, Oxford, United Kingdom}}
\newcommand{\INSTBB}{\affiliation{UPMC, Universit\'e Paris Diderot, CNRS/IN2P3, Laboratoire de Physique Nucl\'eaire et de Hautes Energies (LPNHE), Paris, France}}
\newcommand{\INSTGC}{\affiliation{University of Pittsburgh, Department of Physics and Astronomy, Pittsburgh, Pennsylvania, U.S.A.}}
\newcommand{\INSTFA}{\affiliation{Queen Mary University of London, School of Physics and Astronomy, London, United Kingdom}}
\newcommand{\INSTE}{\affiliation{University of Regina, Department of Physics, Regina, Saskatchewan, Canada}}
\newcommand{\INSTGD}{\affiliation{University of Rochester, Department of Physics and Astronomy, Rochester, New York, U.S.A.}}
\newcommand{\INSTBC}{\affiliation{RWTH Aachen University, III. Physikalisches Institut, Aachen, Germany}}
\newcommand{\INSTFB}{\affiliation{University of Sheffield, Department of Physics and Astronomy, Sheffield, United Kingdom}}
\newcommand{\INSTDI}{\affiliation{University of Silesia, Institute of Physics, Katowice, Poland}}
\newcommand{\INSTEH}{\affiliation{STFC, Rutherford Appleton Laboratory, Harwell Oxford,  and  Daresbury Laboratory, Warrington, United Kingdom}}
\newcommand{\INSTCH}{\affiliation{University of Tokyo, Department of Physics, Tokyo, Japan}}
\newcommand{\INSTBJ}{\affiliation{University of Tokyo, Institute for Cosmic Ray Research, Kamioka Observatory, Kamioka, Japan}}
\newcommand{\INSTCG}{\affiliation{University of Tokyo, Institute for Cosmic Ray Research, Research Center for Cosmic Neutrinos, Kashiwa, Japan}}
\newcommand{\INSTGI}{\affiliation{Tokyo Metropolitan University, Department of Physics, Tokyo, Japan}}
\newcommand{\INSTF}{\affiliation{University of Toronto, Department of Physics, Toronto, Ontario, Canada}}
\newcommand{\INSTB}{\affiliation{TRIUMF, Vancouver, British Columbia, Canada}}
\newcommand{\INSTG}{\affiliation{University of Victoria, Department of Physics and Astronomy, Victoria, British Columbia, Canada}}
\newcommand{\INSTDJ}{\affiliation{University of Warsaw, Faculty of Physics, Warsaw, Poland}}
\newcommand{\INSTDH}{\affiliation{Warsaw University of Technology, Institute of Radioelectronics, Warsaw, Poland}}
\newcommand{\INSTFD}{\affiliation{University of Warwick, Department of Physics, Coventry, United Kingdom}}
\newcommand{\INSTGE}{\affiliation{University of Washington, Department of Physics, Seattle, Washington, U.S.A.}}
\newcommand{\INSTGH}{\affiliation{University of Winnipeg, Department of Physics, Winnipeg, Manitoba, Canada}}
\newcommand{\INSTEA}{\affiliation{Wroclaw University, Faculty of Physics and Astronomy, Wroclaw, Poland}}
\newcommand{\INSTH}{\affiliation{York University, Department of Physics and Astronomy, Toronto, Ontario, Canada}}

\INSTC
\INSTEE
\INSTFE
\INSTD
\INSTGA
\INSTI
\INSTGB
\INSTFG
\INSTFH
\INSTBA
\INSTEF
\INSTEG
\INSTDG
\INSTCB
\INSTED
\INSTEC
\INSTEI
\INSTGF
\INSTBE
\INSTBF
\INSTBD
\INSTEB
\INSTHA
\INSTCC
\INSTCD
\INSTEJ
\INSTFC
\INSTFI
\INSTJ
\INSTCE
\INSTDF
\INSTFJ
\INSTGJ
\INSTCF
\INSTGG
\INSTBB
\INSTGC
\INSTFA
\INSTE
\INSTGD
\INSTBC
\INSTFB
\INSTDI
\INSTEH
\INSTCH
\INSTBJ
\INSTCG
\INSTGI
\INSTF
\INSTB
\INSTG
\INSTDJ
\INSTDH
\INSTFD
\INSTGE
\INSTGH
\INSTEA
\INSTH

\author{K.\,Abe}\INSTBJ
\author{J.\,Adam}\INSTFJ
\author{H.\,Aihara}\INSTCH\INSTHA
\author{T.\,Akiri}\INSTFH
\author{C.\,Andreopoulos}\INSTEH
\author{S.\,Aoki}\INSTCC
\author{A.\,Ariga}\INSTEE
\author{T.\,Ariga}\INSTEE
\author{S.\,Assylbekov}\INSTFG
\author{D.\,Autiero}\INSTJ
% ...

\collaboration{The T2K Collaboration}\noaffiliation

%\collaboration{The T2K Collaboration}%\noaffiliation

\date{\today}% It is always \today, today,

\begin{abstract}
The T2K experiment has observed electron neutrino appearance in a muon neutrino beam produced 295 km from the Super-Kamiokande detector with a peak energy of 0.6 GeV. A total of 28 electron neutrino events were detected with an energy distribution consistent with an appearance signal, corresponding to a significance of 7.3$\sigma$ when compared to 4.92 $\pm$ 0.55 expected background events.
In the PMNS mixing model, the electron neutrino appearance signal depends on several parameters including three mixing angles $\theta_{12}$,  $\theta_{23}$, $\theta_{13}$,  a mass difference $\Delta m^2_{32}$ and a CP violating phase $\delta_{\mathrm{CP}}$.
In this neutrino oscillation scenario, assuming $|\Delta m^2_{32}| = 2.4 \times 10^{-3}$~$\rm eV^2$, $\sin^2 \theta_{23} = 0.5$, and  $\Delta m^2_{32} >0$ ($\Delta m^2_{32} <0$), a best-fit value of $\sin^2 2 \theta_{13}$ = $0.140^{+0.038}_{-0.032}$ ($0.170^{+0.045}_{-0.037}$) is obtained at 
$\delta_{\mathrm{CP}}=0$. When combining the result with the current best knowledge of oscillation parameters including 
the world average value of $\theta_{13}$ from reactor experiments, some values of $\delta_{\mathrm{CP}}$ are disfavored
at the 90\% CL.

% Many ASG and EC members do not want to mention delta_CP in the abstract.
%Using constraints on the three mixing angles and mass differences from other experiments and T2K measurements, a study of the preferred $\delta_{\mathrm{CP}}$ phase space is presented.

%The T2K experiment has observed electron neutrino appearance in a muon neutrino beam produced 295 km from 
%the Super-Kamiokande detector with a peak energy of 0.6 GeV. 
%We observe 28 electron neutrino events, which corresponds to a significance of 7.5$\sigma$ when compared to 4.64 $\pm$ 0.53 background events. 
%In a neutrino oscillation scenario with $|\Delta m^2_{32}| = 2.4 \times 10^{-3}$~$\rm eV^2$, $\theta_{23} = \pi/4$, $\delta_{\mathrm{CP}}=0$, and 
%$\Delta m^2_{32} >0$ ($\Delta m^2_{32} <0$), 
%the best fit value of $\sin^2 2 \theta_{13}$ is $0.150^{+0.039}_{-0.034}$ ($0.182^{+0.046}_{-0.040}$).
%T2K also excludes $\delta_{\mathrm{CP}}$ values between $0.22 \pi$ and $0.84 \pi$ radians at 90\% confidence level.
%This measurement excludes $\delta_{\mathrm{CP}}$ values between $0.22 \pi$ and $0.84 \pi$ ($0.09 \pi$ and $3.1 \pi$) radians
%assuming normal (inverted) hierarchy at 90\% CL.

%This measurement excludes $\delta_{\mathrm{CP}}$ values between $0.22 \pi$ and $0.84 \pi$ ($0.09 \pi$ and $3.1 \pi$) radians
%assuming normal (inverted) hierarchy at 90\% CL.

\end{abstract}
\maketitle

%{\it Introduction}\textemdash 
%An update on the $\nu_\mu \to \nu_e$ oscillation measurement from the T2K experiment~\cite{Abe:2011ks,Abe:2011sj,Abe:2013xua}
%that demonstrates, at the 7.5$\sigma$ level, evidence for $\nu_e$ appearance is presented. 
{\it Introduction}\textemdash 
The discovery of neutrino oscillations using atmospheric neutrinos was made by Super-Kamiokande in 1998~\cite{Fukuda:1998mi}. Since then, many other experiments have confirmed the phenomenon of neutrino oscillations through various disappearance modes of flavor transformations. 
% TN: 2013/11/19  modified based on comments by Hiro
%However, to date, there has not been a definitive observation of neutrino oscillations through the explicit appearance of a neutrino flavor from another neutrino flavor. 
%
However, to date, there has not been an observation of the explicit appearance of a different neutrino flavor from neutrinos of another flavor through neutrino oscillations.
In 2011, the T2K collaboration published the first indication of electron neutrino appearance from a muon neutrino beam at 2.5$\sigma$ significance based on a data set corresponding to $1.43\times 10^{20}$ protons on target (POT)~\cite{Abe:2011ks,Abe:2011sj}. This result was followed by the publication of 
further evidence for electron neutrino appearance at 3.1$\sigma$ in early 2013~\cite{Abe:2013xua}. 
This paper presents new results from the T2K experiment that establish, at greater than $5\sigma$, the observation of electron-neutrino appearance from a muon-neutrino beam.

In a three-flavor framework, neutrino oscillations are described by the PMNS matrix~\cite{Maki:1962mu, Pontecorvo:1967fh} 
which is parameterized by three mixing angles $\theta_{12}$,  $\theta_{23}$, $\theta_{13}$, and a CP
violating phase $\delta_{\mathrm{CP}}$. 
In this framework the probability for $\nu_\mu \to \nu_e$ oscillation can be expressed~\cite{Freund:2001pn} as
%
%\begin{linenomath}
%\begin{align}
%
\begin{small}
\begin{eqnarray}
\lefteqn{P(\nu_\mu \rightarrow \nu_e) \simeq 
\sin^2\theta_{23} \sin^22\theta_{13} \sin^2 \frac{\Delta m^2_{31}L}{4E} }  \notag \\  
& - &  \frac{\sin 2 \theta_{12} \sin 2 \theta_{23}}{2 \sin \theta_{13}}  \sin \frac{\Delta m^2_{21}L}{4E} 
\sin^2 2\theta_{13}  \sin^2 \frac{\Delta m^2_{31}L}{4E} \sin \delta_{\mathrm{CP}} \notag \\ 
& + &  {\rm (CP~even~term,~solar~term,~matter~effect~term),} 
\label{eq:oscprob} 
\end{eqnarray}
\end{small}
%\end{align}
%\end{linenomath}
\hspace{-0.15cm}where $L$ is the neutrino propagation distance and $E$ is the neutrino energy. 
%
%
%The angle $\theta_{13}$ is the dominant parameter governing $\nu_\mu \to \nu_e$ oscillations, 
%while the CP violation contribution appears in sub-leading terms.% that contain $\delta_{\mathrm{CP}}$.
%
The measurement of $\nu_\mu \to \nu_e$ oscillations is of particular
interest because this mode is sensitive to both $\theta_{13}$ and $\delta_{\mathrm{CP}}$. 
The first indication of non-zero $\theta_{13}$ was published by T2K~\cite{Abe:2011sj} 
based on the measurement of $\nu_\mu \to \nu_e$ oscillations. 
%based our preliminary measurement of $\nu_\mu$ to $\nu_e$ oscillations. 
More recently, indications of $\nu_\mu \to \nu_e$ oscillations were also reported by the MINOS experiment~\cite{Adamson:2013ue}.
The value of $\theta_{13}$ is now precisely known to be $9.1\pm0.6^{\circ}$ from measurements of $\overline{\nu}_e$ disappearance in reactor neutrino experiments~\cite{Beringer:1900zz, An:2012eh, Ahn:2012nd, Abe:2011fz}. 
Using the reactor measurement of $\theta_{13}$, the $\nu_\mu \to \nu_e$ appearance mode
can be used to explore CP violation, which has yet to be observed in the lepton sector.
CP violation, as shown in Equation~\ref{eq:oscprob}, is governed by the second term   
and can be as large as 27\% of the first term for the T2K experimental setup when using current values of the neutrino oscillation parameters.

% in particle physics.

% Therefore, by combining the $\nu_\mu$ to $\nu_e$
%measurement with recent reactor $\theta_{13}$ measurements,
%$\delta_{CP}$ can be determined.
%
%Recently, T2K has reported a measurement of $\nu_\mu$ to $\nu_e$ oscillation using data accumulated with
%3.01$\times10^{20}$ protons on target, and MINOS also reported the
%measurement of $\nu_e$ and anti-$\nu_e$ appearance. 
%

{\it T2K Experiment}\textemdash 
T2K operates at the J-PARC facility in Tokai, Japan. 
A muon neutrino beam is produced from the decay of charged pions and kaons generated by 30~GeV protons hitting a graphite target and focused by three magnetic horns.
Downstream of the horns is the decay volume, 96 meters in length, followed by the beam dump and muon monitors~(MUMON).
The neutrino beam illuminates an on-axis detector and off-axis detectors positioned at an angle of 2.5$^\circ$ relative to the beam direction. The resulting
energy spectrum, peaked at 0.6 GeV for the off-axis detectors, reduces the $\nu_e$ contamination and the feed-down backgrounds to the $\nu_e$ appearance signal from higher energy neutrinos. The near detector complex at 280 meters from the target is used to measure the neutrino beam direction, spectrum, and composition before oscillations and to measure neutrino cross sections.
%The complex consists of an on-axis detector (INGRID) and a set of off-axis detectors (P0D-$\pi0$ Detector, FGD-Fine Grain Detector, TPC, ECAL and SMRD-Side Muon Range Detector; collectively, ND280); the details are described in~\cite{Abe:2011ks}.
%The complex consists of an on-axis detector and a suite of off-axis detectors collectively known as ND280. The details of ND280 are described in~\cite{Abe:2011ks}.
The complex consists of an on-axis detector~(INGRID) and a suite of off-axis detectors~(ND280) that reside within a 0.2~T magnet~\cite{Abe:2011ks}.
The Super-Kamiokande (SK) 50~kt water Cherenkov  detector,  situated 295 km away, is used to detect the oscillated neutrinos.
%after oscillation.
 
The results presented here are based on data taken from January 2010 to May 2013.
%
% four running periods: Run 1 (January-June 2010), Run 2 (November 2010-March 2011), Run 3 (January-June 2012) and Run 4 (October 2012-May 2013). 
%The Run 2 period was terminated by the March 11, 2011 Tohoku earthquake, but the accelerator and detector resumed data-taking operations in January 2012. 
During this period the proton beam power has steadily increased and reached 220 kW continuous operation with a world record of $1.2 \times 10^{14}$ protons per pulse.  The total neutrino beam exposure at SK corresponds to $6.57 \times 10^{20}$~POT.

%%%%%%%%%%%%%%%
% Beam flux prediction 
%%%%%%%%%%%%%%
{\it Neutrino Beam Flux}\textemdash 
The neutrino beam flux~\cite{Abe:2012av}  %[Ref. flux paper] 
is predicted by modeling interactions of the primary beam protons 
in a graphite target using external hadron production data from the
CERN NA61/SHINE experiment~\cite{Abgrall:2011ae, Abgrall:2011ts} %[ref. 11, 12 of flux paper]
and the FLUKA2008 package~\cite{FLUKA,FLUKA08}. %[ref. 22,21 of flux paper] 
GEANT3~\cite{GEANT3}       %[ref. 49 of nue paper] 
with GCALOR~\cite{Zeitnitz:1992vw} %[ref. 50 of nue paper] 
simulates propagation of the secondary/tertiary pions and kaons, 
and their decays into neutrinos.
The $\nu_e$ component (including a small amount of $\overline{\nu}_e$) 
in the beam is estimated to be less than 1\% of the flux below 1.5 GeV, 
and constitutes an irreducible background to the $\nu_e$ appearance search. 
This component is generated predominantly  by the decay of muons for $E_\nu<$ 1 GeV
and by kaons for $E_\nu>$ 1 GeV.
 
The neutrino flux uncertainties are dominated 
by the hadron production uncertainties, 
with contributions from the neutrino beam direction 
and the proton beam uncertainties.
The neutrino beam direction, monitored 
indirectly by MUMON on a spill-by-spill basis, 
and directly by INGRID~\cite{Abe:2011xv}, was found to be well 
within the required $\pm$1~mrad
during the full run period. 
INGRID also measured the neutrino interaction rate per POT 
to be stable within 0.7\%. 
The total systematic error for the absolute flux prediction 
is evaluated to be 10--15\% in the relevant energy range.
Furthermore, the uncertainty on the ratio of the flux predictions 
at the far and the near detectors is 
less than 2\% around the peak.

%%%%%%%%% Cross Sections %%%%%%%%%%%%%%%%
{\it Neutrino Interaction Simulations and Cross Section Parameters}\textemdash
The NEUT neutrino interaction generator~\cite{Hayato:2009} is used to simulate neutrino interactions in the INGRID, ND280, and SK detectors. At interaction energies typical of the T2K beam, the dominant charged current (CC) interactions are charged current quasi-elastic (CCQE) and single resonant pion production. The cross section parameterization can be divided into two categories: parameters common to interactions at both ND280 and SK, and parameters evaluated separately for the two detectors. Parameters in the first category comprise the axial masses for CCQE ($M_{A}^{QE}$) and single resonant pion production ($M_{A}^{RES}$), and normalizations for CCQE, CC single pion, and neutral current (NC) 1$\pi^0$ interactions. Parameters in the second category are typically related to the interaction target---primarily carbon at ND280 and oxygen at SK---and include Fermi momentum, binding energy, and spectral function modeling for the CCQE nuclear model. Also in this category are normalizations for other CC and NC cross sections,  the $\nu_{e}/\nu_{\mu}$ CC cross section ratio,
%the ratio $\sigma_{\nu_eCC}/\sigma_{\nu_{\mu}CC}$,
pion production parameters, and final state interactions of pions exiting the nucleus. External data sets, primarily from~\cite{mb-cc1pi0,mb-cc1pip,mb-nc1pi0}, are used to determine the initial values and prior uncertainties of the parameters~\cite{Abe:2013xua}.

{\it ND280 Measurements, Flux and Common Cross Section fits}\textemdash 
The energy spectrum of the neutrino beam and the neutrino cross
section parameters are constrained using $\nu_{\mu}$ CC interactions
in ND280. The fine-grained detectors (FGDs)~\cite{Amaudruz:2012pe} are scintillator trackers
that serve as the primary neutrino target, and the momentum and identity of
the particles emerging from the interaction are determined by the time
projection chambers (TPCs)~\cite{Abgrall:2010hi} interleaved with the FGDs. The muon is assumed to be the
highest-momentum, negative-curvature track that emerges from the FGD
fiducial volume with an energy deposition consistent with a muon in
the TPC downstream of the FGD.  Tracks found in the TPC upstream of
the FGD are used to veto external background events.
%
%The previous analysis divided the CC sample into a "CCQE-like" sample
%and a "CCnonQE-like" sample. The CCQE-like sample was the subset of
%the CC sample which contained only a single TPC-FGD track and no late
%time energy deposit in FGD1 consistent with an electron from a
%$\pi^+\to\mu^+\to e^+$ decay chain. All other CC events were placed in
%the CCnonQE-like sample. These two samples were then simultaneously
%fit to the data.
%
%With the addition of the Run 3 data, the near detector analysis was
%limited by systematic uncertainties. Run 3 increased the number of
%protons on target for which near detector data were collected from
%0.96$\times 10^{20}$ to 2.56$\times 10^{20}$, however these additional
%data only provided a reduction in the uncertainty in the far
%detector event rate prediction from 5.7\% to 4.7\% for
%sin$^22\theta_{13}$ of 0.1. In particular, the fit had limited ability
%to constrain the cross section parameters associated with CC$\pi^+$
%interactions, since the CCnonQE-like sample had a CC$\pi^+$ purity of
%only 29\%. In addition, discrepancies between the neutrino interaction
%model and the data at very forward muon angles were not adequately
%resolved due to the coarseness of the binning.
%
%The  selected with the same criteria as the previous analysis

The ND280 analysis includes many improvements over the previous T2K
electron neutrino appearance measurement~\cite{Abe:2013xua}.
Candidate events are now divided into three samples: CC-0$\pi$,
dominated by CCQE interactions; CC-1$\pi^+$, dominated by CC resonant
pion production; and CC-other.  The samples are defined by the number
of pions in the observed final state.  A $\pi^+$ can be identified in
one of three ways: an FGD+TPC track with positive curvature and a 
TPC charge deposition consistent with a pion,
 an FGD-contained track with a charge
deposition consistent with a pion, or a delayed energy deposit due to
a decay electron from stopped $\pi^+ \to \mu^+$ in the FGD.  To tag a
$\pi^-$, only negative curvature FGD+TPC tracks are used.  A $\pi^0$
is identified if there exists a track in the TPC with a charge
deposition consistent with an electron from a $\gamma$ conversion.
Events containing no pions are classified as CC-0$\pi$, events with
exactly one $\pi^+$ and no $\pi^-$ or $\pi^0$ are classified as
CC-1$\pi^+$, and all other CC events are classified as CC-other.  There
are 17369, 4047, and 4173 data events in the
CC-0$\pi$, CC-1$\pi^+$, and CC-other samples, respectively. The ND280
data set used for this analysis corresponds to
$5.90\times 10^{20}$ POT.

%When the near detector constraint is extracted using the same Run 1-3
%data as the previous analysis, but with the new CC0$\pi$, CC1$\pi^+$,
%and CCother samples, the uncertainty on the far detector event rate is
%reduced from 4.7\% to 3.5\% for sin$^22\theta_{13}$ of 0.1,
%sin$^22\theta_{23}$ of 1.0, and $\delta_{CP}$ of 0.

\begin{figure}[tbp]
 \vspace{3mm}
 \includegraphics[width=0.45\textwidth]{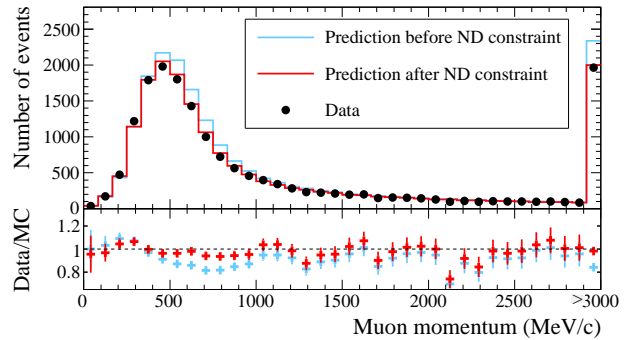} 
 \caption{The muon momentum distribution for the ND280 CC-0$\pi$ sample (upper). 
  The black points represent the data, the blue histogram shows the MC prediction before
  data constraint, and the red histogram shows the MC prediction after constraint. The lower plot shows the ratio of data to MC for the pre- and post-constraint cases.}  \label{fig:ND280momentumCCQE}
\end{figure}

The three samples are fit with 25 beam flux parameters at ND280 (11
$E_{\nu_{\mu}}$, 5 $E_{\bar{\nu}_{\mu}}$, 7 $E_{\nu_{e}}$, and 2
$E_{\bar{\nu}_{e}}$ bins), 21 cross section parameters (5 in common
with SK, and 16 used only for ND280), as well as 210 parameters describing
the ND280 detector systematics (10 momentum
$\times$ 7 angle bins for each sample). The dominant detector uncertainties come 
from events occurring outside the FGD fiducial volume and from pion reinteractions in
 the detector. The ND280 measurements constrain the SK flux parameters due to the flux covariance derived from beam simulations.
 The predicted numbers of ND280 events in Monte Carlo (MC), 
 using the best-fit parameters, are 17352, 4110, and 4119 
 for the CC-0$\pi$, CC-1$\pi^+$, and CC-other samples, respectively.
 A $\chi^2$ goodness-of-fit test returns a p-value of 0.66, indicating no disagreement between
 the data and the prediction using best-fit parameters.
 %Table~\ref{tab:ND280rates} shows the number of observed events in ND280 for each sample with 
%the numbers predicted in the Monte Carlo simulation (MC) prior to the fit and after the fit. 
Figure~\ref{fig:ND280momentumCCQE} shows the muon momentum distribution
of the CC-0$\pi$ sample, and the improvement in data and MC agreement when using the best-fit parameters. %In all three cases, the agreement between data and MC is significantly improved by the constraint.
%described previously. 

The fit to the ND280 data gives estimates for 22 beam flux parameters at SK, the 5 common cross section parameters, and their covariance. Using the ND280 information reduces the uncertainty on the expected number of electron-like events at SK due to the propagated parameters from 25.9\% to 2.9\%.

%\begin{table}[ht]
%\begin{center}
%\caption{The number of observed events in ND280 for each sample with 
%the numbers predicted in the Monte Carlo simulation (MC) prior to the fit and after the fit.}
%\label{tab:ND280rates}
%\begin{tabular}{lccc}
%\hline\hline
%                      &  CC-0$\pi$ & CC-1$\pi^+$  & CC-Other    
%\\  \hline
%Data                  &  16912    & 3936       & 4062       \\
%MC (prior to fit)  &  20016.2  & 5059.4    & 4602.1        \\ 
%MC (after fit) &  16802.5  & 3970.3    & 4006.0         \\ \hline\hline
%\end{tabular}
%\end{center}
%\end{table}

{\it SK Measurements}\textemdash 
The SK detector is composed of an inner detector (ID) and an outer detector (OD). The ID has a water fiducial volume (FV) of 22.5 kt that is equipped with 11129 photomultiplier tubes (PMT) and is surrounded by the 2~m wide OD.
%The SK detector has a fiducial volume (FV) of 22.5~kt in its inner detector (ID), which is equipped with 11129 photomultiplier tubes (PMTs), and surrounded by a 2~m wide outer detector (OD).
%The 2~m-wide outer detector (OD) completely encloses the ID.
Neutrino events at SK are selected if the Cherenkov ring is consistent with an
energy above 30~MeV in the ID with low activity in the OD to reject any entering background or exiting events.  
These events are labeled fully-contained (FC). 
The FC fiducial volume (FCFV) sample is obtained by applying the further cut that the event vertex is at least 2~m away from the ID tank wall.
%By applying the additional cut that the event vertex is at least 2 m away from the ID tank wall, the events are then further categorized as the FC fiducial volume (FCFV) sample.
A timing cut of $-2$ to 10 $\mu$s relative to the first beam bunch arrival is applied to distinguish T2K data from other neutrino samples such as
atmospheric neutrino interactions.
%A clear peak with a 24 ns width in RMS of the expected arrival time can be seen.  
The timing cut reduces the contamination from other neutrino sources to 0.0085 events in the full sample.   

To select $\nu_{e}$ interaction candidate events in the FCFV sample,
a single electron-like Cherenkov ring is required.
%, which rejects 97.7\% of the $\nu_{\mu}$ and $\bar{\nu}_{\mu}$ CC events.  
The reconstructed electron momentum ($p_e$) is required to exceed 100 MeV/c to eliminate decay-electrons from stopping muons generated by CC interactions
and pions in NC interactions. 
In addition, events are required to have a reconstructed neutrino energy ($E_{\nu}^{\rm rec}$) below 1250~MeV. %, as $\nu_{e}$ events below this are mostly oscillated events and above this are mostly intrinsic beam events. 
Nearly all of the oscillated $\nu_{e}$ signal events are below this value, 
while most of the intrinsic beam $\nu_e$ background events have higher energies. 
The $E_{\nu}^{\rm rec}$ is calculated assuming a CCQE interaction as

\noindent
\begin{equation}
E_{\nu}^{\rm rec} = \frac{
m_{p}^{2} - (m_{n} - E_{b})^{2} - m_{e}^{2} + 2(m_{n}-E_{b}) E_e}
{2(m_{n}-E_{b}-E_e+p_e \cos\theta_{\rm e})} ,   \label{eq:ccqe}
\end{equation}
where $m_{n}$ ($m_{p}$) is the neutron (proton) mass, $E_{b}$ is the
neutron binding energy in oxygen (27 MeV), $m_e$ is the electron mass, 
$E_e$ is its energy,  and $\theta_{\rm e}$ is the angle of the electron direction relative to the 
beam direction.

\begin{figure}[h] \begin{center}
    \begin{center}
      \begin{minipage}[t]{.45\textwidth}
        \begin{center}
          \includegraphics[width=\textwidth] {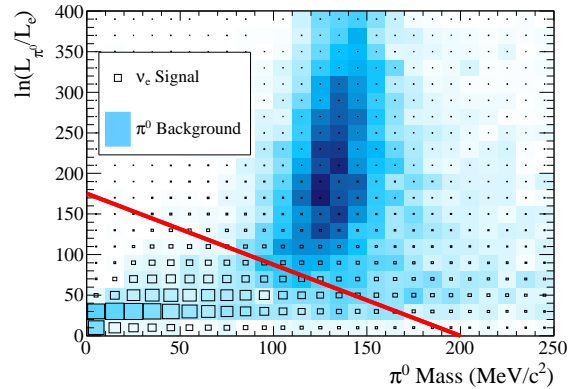}
        \end{center}
      \end{minipage}
    \end{center}
 \caption{The $\ln(L_{\pi^0}/L_e)$ vs $m_{\pi^0}$ distribution is shown for both signal $\nu_e$-CC events (boxes) and background events containing a $\pi^0$ (blue scale). 
The red line indicates the location of the $\pi^0$ rejection cut. Events in the upper right corner are rejected.}
\label{fig:fqpi0cut}
\end{center}
\end{figure}

The final selection criterion removes additional $\pi^0$ background
events using a new reconstruction algorithm, based on an extension of
the model described in Reference~\cite{miniboonerecon}, to determine
the kinematics of all final state particles. The new algorithm is a
maximum-likelihood fit in which charge and time probability density
functions (PDFs) are constructed for every PMT hit for a given
particle hypothesis with a set of 7 parameters:
%the 4-vertex ($x$, $y$, $z$, and $t$), the direction ($\theta$ and $\phi$), and the momentum. 
the vertex position, the timing, the direction and the momentum.
Multiple-particle fit hypotheses are constructed by summing the charge
contributions from each constituent particle.  Different neutrino final states
are distinguished by comparing the best-fit likelihood
resulting from the fit of each hypothesis. To separate $\pi^0$ events
from $\nu_e$ CC events, both the reconstructed $\pi^0$ mass
($m_{\pi^0}$) and the ratio of the best-fit likelihoods of the $\pi^0$
and electron fits ($L_{\pi^0}/L_{e}$) are used.
Figure~\ref{fig:fqpi0cut} shows the $\ln(L_{\pi^0}/L_{e})$ vs $\pi^0$
mass distribution for signal $\nu_e$-CC events and events containing a
$\pi^0$ in the MC sample, as well as the rejection cut line.  Events
that satisfy $\ln(L_{\pi^0}/L_{e})<175-0.875 \times
m_{\pi^0}$~(MeV/c$^{2}$) constitute the final $\nu_e$ candidate
sample.  This cut removes 69\% of the $\pi^0$ background events
relative to the previous T2K $\nu_e$ appearance selection, with only a
2\% loss in signal efficiency~\cite{Abe:2011sj}.

A summary of the number of events passing each selection cut is shown in Table~\ref{tab:skreduction}. 
After all cuts, the total number of candidate $\nu_e$ events selected in data is 28, which 
%, and the prediction from the simulation for
%$\textrm{sin}^22\theta_{13}=0.1$, $\textrm{sin}^22\theta_{23}=1$, and $\delta_{CP}=0$ is
%20.41 (updated to final BANFF-weighted number).
%The 28 $\nu_{\mu} \rightarrow \nu_{e}$ events 
is significantly larger than the 4.92$\pm$0.55 %(syst.) 
expected events for $\theta_{13} = 0$. 
For $\textrm{sin}^22\theta_{13}=0.1$ and $\delta_{\mathrm{CP}}=0$, the expected number is 21.6, as shown in Table~\ref{tab:skreduction}. 
%20.41 (updated to final BANFF-weighted number).

\begin{table}[th]
\begin{center}
\caption{
%\textcolor{red}{This table probably has too much information for a PRL, but I included it here for completeness, and to allow comments about what to remove} \textcolor{black}
The expected number of signal and background events passing each selection stage assuming
$\textrm{sin}^{2}2\theta_{13}=0.1$, $\sin^2 \theta_{23}=0.5$, $|\Delta m^2_{32}|=2.4\times10^{-3}$ eV$^{2}$, $\delta_{\mathrm{CP}}=0$, and $\Delta m^2_{32}>0$, compared to the observed number in data. Interactions in the true FV are based on the MC truth information while all other numbers are based on the reconstructed information and have been rounded off after addition to avoid rounding error.
%Expectations are normalized to $6.393\times10^{20}$POT.
} \label{tab:skreduction}
\begin{tabular}{lc|ccccc}
\hline
\hline
Selection                      & \multirow{2}{*}{Data}   & \ensuremath{\nu_\mu}$\rightarrow$\nue    & \ensuremath{\nu_\mu}+\numb & \ensuremath{\nu_e}+\nueb & \multirow{2}{*}{NC}    & Total  \\ % & BG    
                          &       & CC  & CC    & CC         &                      &   MC  \\ % & total 
\hline                                                                                                  %         
 Interactions in FV                    & -  &  27.1  & 325.7 & 16.0  & 288.1   & 656.8  \\ \hline % & 637.02
% \hspace{4pt}                        &           &           &       &              &  \\ \hline
%\hline                                                                                                  %         
 FCFV                                  & 377  & 26.2 & 247.8     & 15.4      & 83.0          &  372.4   \\ % & 358.96
+Single-ring                           & 193  & 22.7 & 142.4     & 9.8      & 23.5          &    198.4\\ % & 181.79
+$e$-like PID                           & 60  & 22.4 & 5.6       & 9.7      & 16.3           & 54.2    \\ % & 33.87 
+$p_e$$>$$100{\rm MeV/c}$                 & 57  & 22.0 & 3.7        & 9.7      & 14.0     &   49.4       \\ % & 29.32 
+No decay-$e$                           & 44  & 19.6 & 0.7  & 7.9       & 11.8           &    40.0 \\ % & 21.94 
+$E_{\nu}^{\rm rec}$$<$$1250{\rm MeV}$    & 39   & 18.8 & 0.2 &  3.7       & 9.0  & 31.7            \\ % & 13.69 
+Non-$\pi^0$-like                      & 28   & 17.3 & 0.1 & 3.2       & 1.0          &    21.6 \\ % & 4.49  
%\hline
%Efficiency [\%]                    & 0.0        & 19.2       & 0.4    & 61.2                  & - \\ % & 0.7
%Efficiency [\%]                       & 0.0        & 19.1       & 0.4   & 61.2  & 3.3         & - \\ % & 1.3 from FCFV 
\hline
\hline
\end{tabular}
\end{center}
\end{table}

The systematic uncertainty due to the SK selection cuts is evaluated using various data and
MC samples.
% by combining all cut systematics in each interaction type. 
The uncertainty for both the FC and the FV selection
%, determined using cosmic ray muons, 
is 1\%.  
The decay-electron rejection cut has errors
of 0.2-0.4\%, depending on neutrino flavor and interaction type.
%For the energy cut, 2.4\% uncertainty was evaluated. 
The uncertainties for the single electron-like ring selection and $\pi^{0}$
rejection are estimated by using the SK atmospheric neutrino data and
SK cosmic-ray muons.  Electron-neutrino CC-enriched control samples
based on these cuts were prepared, and the differences between MC
predictions and data are used to extract the systematic uncertainty.
The uncertainty associated with the $\pi^{0}$ background is determined by
constructing a hybrid sample with either an electron-like ring taken
from the atmospheric data sample or from decay-electrons selected in
the stopping muon data sample, and a MC-generated gamma ray assuming
$\pi^{0}$ kinematics.
%The error associated with the $\pi^{0}$ background is determined by constructing a hybrid sample of an electron-like ring from the atmospheric data sample and a MC generated gamma ray assuming  $\pi^{0}$ kinematics. 
The selection cut systematic uncertainty is calculated to be 1.6\% for
signal events and 7.3\% for background events.  The total SK selection
uncertainty is 2.1\% for the $\nu_{e}$ candidate events assuming
$\textrm{sin}^{2} 2\theta_{13} = 0.1$.
%The total SK detector systematic error is 2.1%, assuming sin22theta13=0.1.

% TN corrects the description of PN because it looks longer by comparing with other parts.
Additional SK systematic uncertainties are due to final-state interactions (FSI)
of pions that occur inside the target nucleus, as well as secondary
interactions (SI) of pions and photo-nuclear (PN) interactions of
photons that occur outside of the target nucleus. 
% TN 2013/10/28
%The treatment of the FSI and SI uncertainties is unchanged relative to the previous T2K $\nu_e$ appearance analysis~\cite{t2knue}. 
The treatment of the FSI and SI uncertainties is the same as in the previous analysis~\cite{t2knue}. 
For this analysis, a new simulation of PN interactions has been added to the SK MC. 
% TN 2013/10/28
%These interactions produce a set of irreducible $\pi^0$ background events by
%absorbing photons produced in $\pi^0$ decay.  
%The PN effect has the most significant impact near the giant dipole resonance at photon
%energies between 20 and 30~MeV and accounts for 15\% of the total
%$\pi^0$ background in the final $\nu_e$ event sample. The MC
%simulation does not produce any final state particles, including
%secondary photons, when a PN interaction takes place. A systematic
%uncertainty of 100\% is assumed for the normalization of the PN cross
%section to account for any additional light produced in PN
%interactions.
In the final $\nu_e$ event sample, 15\% of the remaining $\pi^0$
background is due to events where one of the $\pi^0$ decay photons is
absorbed in a PN interaction.  A systematic uncertainty of 100\% is
assumed for the normalization of the PN cross section.

%The full set of SK systematic uncertainties is given in Table~\ref{tbl:totalsystematic}.

% to be added to symbols.tex
%%%%%%%%%%%%%%%%% Neutrino Mixing   %%%%%%%%%
\def\stst        {\ensuremath{\textrm{sin}^2(2\theta_{13})}\xspace} 
\def\mdmsq     {\ensuremath{|\Delta m^{2}_{32}|\xspace}}
\def\evsqc    {\ensuremath{\rm \,eV^{2} / c^{4}\xspace}}
%%%%%%%%%% Oscillation fit %%%%%%%%%%%%%%%%%%%%%
\def\sysp        {\ensuremath{\boldsymbol{f}}}
\def\nskexp    {\ensuremath{n_{\textrm{SK}}^{\textrm{exp}}}}
\def\nskobs    {\ensuremath{n_{\textrm{SK}}^{\textrm{obs}}}}
\def\nexp    {\ensuremath{n^{\textrm{exp}}}}
\def\nobs    {\ensuremath{n^{\textrm{obs}}}}
\def\LLnd       {\ensuremath{{\cal L}_{\textrm{ND280}}}} 
\def\LLsk       {\ensuremath{{\cal L}_{\textrm{SK}}}} 
%%%%%%%%%%%%%%%%%% LIGHT MESONS  %%%%%%%%%%%%%%%%%
\def\pion  {\ensuremath{\pi}\xspace}
%%%%%%%%%%%% DISTANCE AND AREA %%%%%%%%%%%%%%%%%%%
\def\cmv  {\ensuremath{{\rm \,cm}^3}\xspace}

{\it Oscillation Analysis}\textemdash 
The neutrino oscillation parameters are evaluated using a binned extended
maximum-likelihood fit.
% using the $\nu_{e}$ appearance candidates. %, as well as $\theta_{23}$ and $\delta_{\mathrm{CP}}$ with $\theta_{23}$e 
The likelihood consists of four components: a normalization term (${\cal
L}_{\mathrm{norm}}$), a term for the spectrum shape (${\cal
L}_{\mathrm{shape}}$), a systematics term (${\cal L}_{\mathrm{syst}}$),
and a constraint term (${\cal L}_{\mathrm{const}}$) from other measurements,
\begin{eqnarray}
\nonumber
{\cal L}(N_{\mathrm{obs}},{\boldmath \vec{x}},
{\boldmath \vec{o}},{\boldmath \vec{f}})
&=&{\cal L}_{\mathrm{norm}}(N_{\mathrm{obs}};{\boldmath \vec{o}},{\boldmath \vec{f}})
 \times{\cal L}_{\mathrm{shape}}({\boldmath \vec{x}}; {\boldmath \vec{o}},{\boldmath \vec{f}})
\\
&& \times{\cal L}_{\mathrm{syst}}({\boldmath \vec{f}})
\times{\cal L}_{\mathrm{const}}({\boldmath \vec{o}}),
\end{eqnarray}
where $N_{obs}$ is the number of observed events, $\vec{x}$ is a set of
kinematic variables, $\vec{o}$ represents oscillation parameters, and
$\vec{f}$ describes systematic uncertainties. 
In the fit, the likelihood is integrated over the nuisance parameters
to obtain a marginalized likelihood for the parameters of interest.

%The normalization term (${\cal L}_{norm}$) corresponds to a likelihood
%for number of $\nu_{e}$ candidate events.
%The ${\cal L}_{\mathrm{norm}}$ is the Poisson probability of reconstructing $N_{\mathrm{obs}}$ events
%using the mean value from the predicted number of MC events.
${\cal L}_{\mathrm{norm}}$ is calculated from a Poisson distribution using the mean value from the predicted number of MC events.
%${\cal L}_{norm}$ is the probability that $N_{obs}$ is drawn from a Poisson distribution with the mean of the predicted number of events. 
${\cal L}_{\mathrm{syst}}({\boldmath \vec{f}})$ constrains the 27 systematic parameters from the ND280 fit, the SK-only cross section parameters, and the SK selection efficiencies.
%constrains the systematic
%uncertainties, which consist of the propagated uncertainties from the ND280 analysis, the SK-only cross section parameters, and the selection uncertainty at SK. %flux, cross section and selection uncertainty at SK. 
%The correlations between the systematic parameters are taken into account. 
%The flux uncertainties and a part of cross section uncertainties are constrained by the prior fit with ND280 data. 
Table~\ref{tbl:totalsystematic} shows the uncertainties on the predicted
number of signal $\nue$ events.
%The total error is significantly reduced 
%from 28.2\% 
%to 8.8\% when including the ND280 constraint.% from ND280 data. 
%The beam flux and the cross section errors which can be constrained by
%using the ND280 data is significantly reduced from 26.4\% to 2.9\%.
%The uncertainty due to beam flux and cross sections is reduced from 26.4\% to 2.9\% by the ND280 data constraint.
%Compared to the previous analysis\cite{}, at $\sin^22\theta_{13}=0.1$,
%the near detector error reduced from 5.0\% to 3.0\%, and the far
%detector error reduced from 3.8\% to 3.5\%, and the total error reduced
%from 9.9\% to 8.8\%.
%
\begin{table}[th]
\begin{center}
\caption{
The uncertainty (RMS/mean in \%) on the predicted number of signal $\nu_e$ events for each group of systematic uncertainties for 
$\textrm{sin}^22\theta_{13}=0.1$ and $0$. The uncorrelated $\nu$ interaction uncertainties are those coming from parts of the neutrino interaction model that cannot be constrained with ND280.
 %Or more detailed explanation is needed?
%The ``Near detector'' category consists of 22 flux parameters and the cross section parameters: $M_{A}^{QE}$ , $M_{A}^{RES}$, normalizations of CC QE, CC 1$\pi$ and NC 1$\pi^{0}$. 
%The rest of 9 cross section parameters, not constrained by ND280, are categorized in ``$\nu$ interaction''.
%The other systematics of SK selection and interactions are categorized in ``Far detector''. 
%The error in bracket shows a systematic uncertainty without a constraint by ND280.
} \label{tbl:totalsystematic}
%\ \\
%Percentage uncertainty in the $\nu_{e}$ candidate events \\% [\%] \\
\begin{tabular}{lcc}
\hline\hline
%\multirow{2}{*}{Error source}
Error source [\%]
%& \multicolumn{2}{c}{$\sin^22\theta_{13}=0.1$}
%& \multicolumn{2}{c}{$\sin^22\theta_{13}=0$} \\
& $\textrm{sin}^22\theta_{13}=0.1$ & $\textrm{sin}^22\theta_{13}=0$ \\ \hline
%RMS/mean [\%] 
%& W/o constraint & Constraint & W/o constraint & Constraint \\ \hline
%Near detector           & 26.9 &  3.0 & 22.2 &  4.9 \\
%$\nu$ interaction       &  7.5 &  7.5 &  6.7 &  6.7 \\
%Far detector            &  3.5 &  3.5 &  7.4 &  7.3 \\ \hline
%Total                   & 28.2 &  8.8 & 24.5 & 11.1 \\ 
Beam flux and near detector           &  2.9   &   4.8  \\
~(w/o ND280 constraint)               & (25.9) & (21.7) \\
Uncorrelated $\nu$ interaction      &  7.5   &   6.8  \\
Far detector and FSI+SI+PN            &  3.5   &   7.3  \\ \hline
Total                                 &  8.8   &  11.1  \\ 
%~(w/o ND280 constraint)        & (28.2) & (24.5) \\ 
%\hline
%Total (2010a analysis) & 20.2 & $^{+17.6}_{-17.5}$ & 25.6 & $^{+22.8}_{-22.7}$ \\
\hline\hline
\end{tabular}
\end{center}
\end{table}
%
%The ${\cal L}_{\mathrm{shape}}$ uses a normalized PDF in ($p_e$, $\theta_e$) bins that represents
%the $\nu_e$ signal and the backgrounds.
The ${\cal L}_{\mathrm{shape}}$ term uses $x$=($p_e$, $\theta_e$) to distinguish 
the $\nu_e$ signal from backgrounds.
%${\cal L}_{shape}$ uses a normalized PDF for the $\nu_{e}$ signal and the backgrounds in 
%electron momentum and angle ($p_e$, $\theta_e$) bins. 
%Electron momentum and angle are the basic observables by which neutrino energy can be calculated in a CC QE interaction as shown in Equation~\ref{eq:ccqe}. 
%
%
%For the background events, it is not the case that the neutrino energy can be reconstructed with ($p_e, \theta_e$).
%The $p_e-\theta_e$ distribution $\nu_e$ appearance signal mainly depends on 
%The kinematics of CC QE interactions, which mainly describes the  $\nu_e$ appearance signal,
%does not describe the background events and 
%therefore signal and background can be distinguished by ${\cal L}_{shape}$. 
%An alternative analysis employs the reconstructed neutrino energy ($E^{rec}_\nu$) distribution as a PDF.
An alternative analysis uses $x=E^{rec}_\nu$, the reconstructed neutrino energy.
%, and the results are consistent with each other. 
%Two analyses confirmed a consistent result independently using an electron momentum and an angle $\vec{x} = (p_e$,$\theta_e)$ or a reconstructed neutrino energy ($\vec{x} = E_{\nu}^{rec}$) in the PDF. 
%%This factor increases the sensitivity of the $\nu_{e}$ appearance analysis because the shapes along the parameters differ between the $\nu_{\mu} \rightarrow \nu_{e}$ signal and backgrounds and can be used to descliminate the signal. 
%Fluxes with different sources and cross section by interactions as a function of the momentum, angle and energy make these difference. 
%
%The maximum likelihood fit is performed to extract the $\theta_{13}$ value 
%For a while, the constraint term, ${\cal L}_{const}$, is not used. 
In order to combine the results presented in this letter with other
measurements to better constrain sin$^22\theta_{13}$ and $\delta_{\mathrm{CP}}$, the
${\cal L}_{\mathrm{const}}$ term can also be used to apply additional constraints on sin$^22\theta_{13}$,
sin$^2\theta_{23}$ and $\Delta m^2_{32}$.
%${\cal L}_{const}$ is used to apply additional constraints on sin$^22\theta_{13}$, sin$^2\theta_{23}$, and
%$\Delta m^2_{32}$ when using results from other published measurements.
%, such as the reactor experiments or T2K measurements.

\begin{figure}[tbp]
\begin{center}
\includegraphics[clip, width=0.5\textwidth]{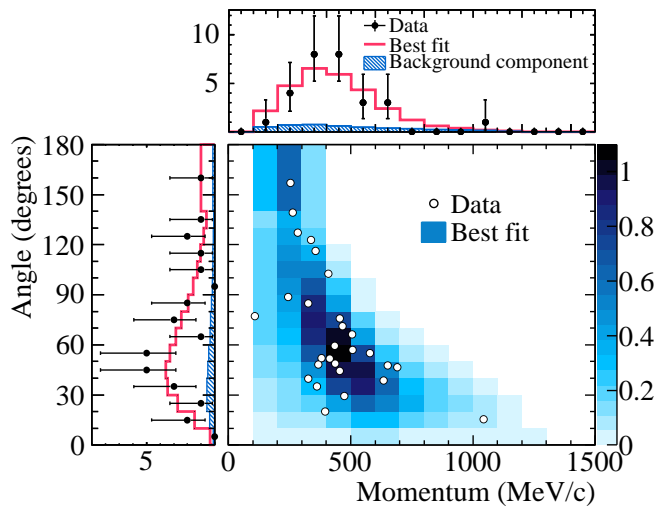}
%   \begin{minipage}{0.4\textwidth}
%     \includegraphics[keepaspectratio=true,width=70mm]{images/real_result_lh21_rn14_mh0_pcosth_st2340}
%   \end{minipage}
%\\
%(a) ($p_e$, $\theta_e$) bin
%\\
%   \begin{minipage}{0.4\textwidth}
%     \includegraphics[keepaspectratio=true,width=70mm]{images/real_result_lh21_rn14_mh0_st2340_pl}
%   \end{minipage}
%\\
%(b) Projection on $p_e$ bin
%\\
%   \begin{minipage}{0.4\textwidth}
%     \includegraphics[keepaspectratio=true,width=70mm]{images/real_result_lh21_rn14_mh0_costh_st2340}
%   \end{minipage}
%\\
%(c) Projection on $\theta_e$ bin
%\\
%   \begin{minipage}{0.4\textwidth}
%     \includegraphics[keepaspectratio=true,width=70mm]{real_result_lh21_rn14_mh0_st2340_dl}
%   \end{minipage}
%   \vspace{10mm}
\caption{The ($p_e$, $\theta_e$) distribution for $\nu_e$ candidate
 events with the MC prediction using the primary method best-fit value of $\textrm{sin}^22\theta_{13} = 0.140$ (normal hierarchy). }
\label{fig:ptheta_bestfit}
\end{center}
\end{figure}

\begin{figure}[tbp]
\begin{center}
\includegraphics[width=0.45\textwidth]{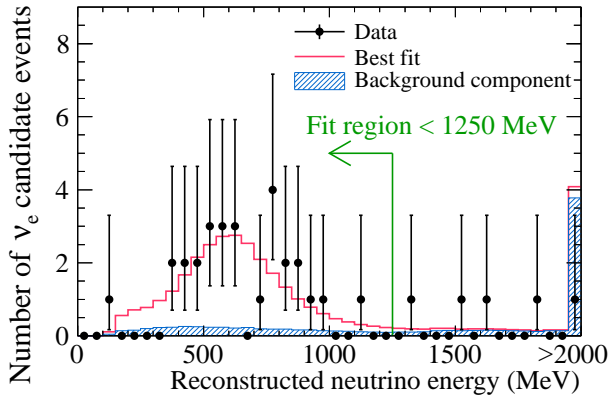}
\caption{The $E_{\nu}^{\rm rec}$ distribution for $\nu_e$ candidate
 events with the MC prediction at the best fit of
 $\textrm{sin}^22\theta_{13} = 0.144$ (normal hierarchy) by the
 alternative binned $E_{\nu}^{\rm rec}$ analysis. }
\label{fig:erec_bestfit}
\end{center}
\end{figure}

%A maximum likelihood fit is performed to extract $\theta_{13}$ while fixing other oscillation parameters: 
The following oscillation parameters are fixed in the analysis:
$\textrm{sin}^{2}\theta_{12} = 0.306$, $\Delta m^{2}_{21} = 7.6 \times
10^{-5}~\mathrm{eV}^{2}$~\cite{Fogli}, $\textrm{sin}^{2}\theta_{23} = 0.5$, 
 $|\Delta m^{2}_{32}| = 2.4 \times 10^{-3}~\mathrm{eV}^{2}$~\cite{Abe:2013fuq} and $\delta_{\mathrm{CP}} = 0$. 
For the normal (inverted) hierarchy case, the best-fit value with a 68\%
confidence level (CL) is  $\textrm{sin}^{2}2\theta_{13} =
0.140^{+0.038}_{-0.032}$ ($0.170^{+0.045}_{-0.037}$). 
Figure~\ref{fig:ptheta_bestfit} shows the best-fit result, with the 28
observed $\nu_{e}$ events. 
%in ($p_e$, $\theta_e$) bins.
%
%
%The alternative analysis using $E_\nu^{rec}$ produced consistent results; the difference between the
%best fit values for the two methods is 0.002 with basically identical
%confidence regions.
%
The alternative analysis using $E_\nu^{rec}$ and a profile likelihood method produces consistent best-fit values and 
%The alternative analysis using $E_\nu^{rec}$ produces consistent best-fit values and 
%the best-fit value of $\textrm{sin}^22\theta_{13} = 0.152$ ($0.184$) with 
nearly identical confidence regions.
Figure~\ref{fig:erec_bestfit} shows the $E_{\nu}^{\rm rec}$ distribution with the MC prediction for the best-fit $\theta_{13}$ value in the alternative analysis.

%It is obtained by the ($p_e$,$\theta_e$)-binned analysis and the
%analysis with $E_{\nu}^{rec}$ bins showed almost consistent values in
%all results. 

The significance for a non-zero $\theta_{13}$ is calculated to be 7.3$\sigma$,
using the difference of log likelihood values between the best-fit  $\theta_{13}$ value and $\theta_{13}=0$. 
%$\theta_{13} \neq 0$. 
%This was determined by using the difference of likelihood values obtained between our best fit  $\theta_{13}$ value and $\theta_{13}=0$.
%($p$-value $= 3.2 \times 10^{-14}$) 
%This result confirms an observation of a $\nu_\mu \to \nu_e$ signal. 
% TN: 2013/11/15 based on Mike's comments
%This result definitely establishes an observation of $\nu_\mu \to \nu_e$ appearance. 
An alternative method of calculating the significance, by generating a large number
 of toy MC experiments assuming $\theta_{13}=0$, also returns a value of 7.3$\sigma$.
%The
%same significance %of 7.3$\sigma$% 
%was also calculated %using a frequentist approach 
%by generating a large number of toy MC experiments assuming $\theta_{13}=0$.
% TN: 2013/11/19 modify with Dean's comment
%Though the significance is calculated at fixed $\delta_{\mathrm{CP}}$ and $\theta_{23}$,
%the significance remains above 7$\sigma$ over all values of $\delta_{\mathrm{CP}}$
%and when $\theta_{23}$ is allowed to vary within its uncertainty.
%
These significances were calculated using a test statistic having fixed values for $\theta_{23}$ and $\delta_{\mathrm{CP}}$. 
For any values for these parameters, consistent with their present uncertainties, the significance remains above 7$\sigma$.

% 
%Whether we change the value of $\theta_{23}$ within its uncertainty or
%change the value of $\delta_{\mathrm{CP}}$, the
%significance do not become lower than 7$\sigma$ for both analyses.
%The significance is greater than 7$\sigma$ for both analyses for all values of $\delta_{\mathrm{CP}}$ and including the uncertainty on $\theta_{23}$.

\begin{figure}[tbp]
\begin{center}
%   \begin{minipage}{0.4\textwidth}
%     \includegraphics[keepaspectratio=true,width=70mm]{images/delta_st13_nh}
%   \end{minipage}
%   \begin{minipage}{0.4\textwidth}
%     \includegraphics[keepaspectratio=true,width=70mm]{images/delta_st13_ih}
%   \end{minipage}
 \includegraphics[width=70mm]{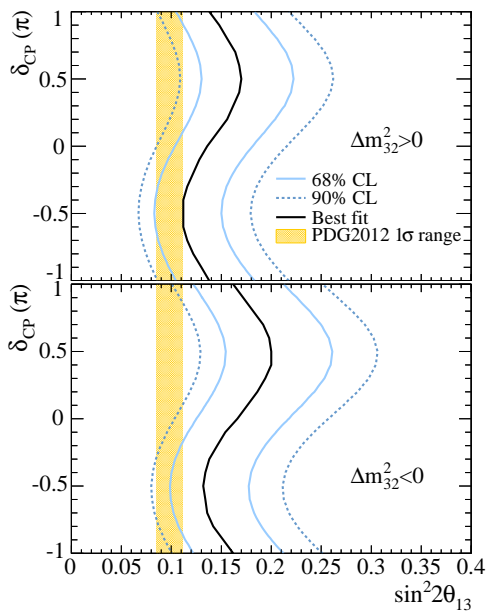}
\caption{
 The 68\% and 90\% CL allowed regions for sin$^22\theta_{13}$, as a
 function %of scanned over values
 of $\delta_{\mathrm{CP}}$ assuming normal hierarchy (top) and inverted
 hierarchy (bottom). The solid line represents the best fit sin$^2 2\theta_{13}$ value
  for given $\delta_{\mathrm{CP}}$  values.
 The values of sin$^2\theta_{23}$ and $\Delta
 m^{2}_{32}$ are varied in the fit with the constraint from \cite{Abe:2013fuq}. 
The shaded region shows the average $\theta_{13}$ value from the
PDG2012~\cite{Beringer:1900zz}.
%The likelihood is marginalized over sin$^2\theta_{23}$
% and $\Delta m^{2}_{32}$.
%The values of sin$^2\theta_{23}$ is fixed to 0.4 (dashed
% line), 0.5 (solid line) and 0.6 (chain line). 
}
\label{fig:delta_st13_contours}
\end{center}
\end{figure}

As the precision of this measurement increases, the uncertainty from
other oscillation parameters becomes increasingly important.
%and this effect can be included in the analysis through the ${\cal
%L}_{const}$ term. 
The uncertainties on $\theta_{23}$ and $\Delta m^2_{32}$ are taken
into account in the fit by adding a ${\cal L}_{\mathrm{const}}$ term 
and marginalizing the likelihood over $\theta_{23}$ and $\Delta m^2_{32}$.
%and integrating the likelihood over $\theta_{23}$ and $\Delta m^2_{32}$
%to obtain the marginalized likelihood. 
The ${\cal L}_{\mathrm{const}}$ term is the likelihood as a function of
sin$^{2}\theta_{23}$ and $\Delta m^2_{32}$, obtained from the T2K $\nu_\mu$
disappearance measurement~\cite{Abe:2013fuq}. 
The value of $\delta_{\mathrm{CP}}$ and the hierarchy are held fixed in the
fit. Performing the fit for all values of $\delta_{\mathrm{CP}}$, the
allowed 68\% and 90\% CL regions for sin$^22\theta_{13}$ are obtained as
shown in Figure~\ref{fig:delta_st13_contours}.
%Figure \ref{fig:delta_st13_contours} shows the allowed 68\% and 90\% CL regions 
%for sin$^22\theta_{13}$, scanned over all values of
%$\delta_{\mathrm{CP}}$ using the constraints on $\theta_{23}$ and
%$\Delta m^2_{32}$ from the T2K $\nu_\mu$ disappearance
%measurement~\cite{Abe:2013fuq}.
%
%Although the value of sin$^2\theta_{23}$ was fixed to 0.5 and is consistent with the T2K disappearance measurement,
%%the uncertainty of sin$^2\theta_{23}$ is take into account when we discuss the constraint on $\delta_\mathrm{CP}$.
%the uncertainty in $\theta_{23}$ and $\Delta m^{2}_{32}$ by the T2K
%$\nu_\mu$ measurement was taken into account in ${\cal L}_{const}$.
%
For $\delta_{\mathrm{CP}}=0$ and normal (inverted) hierarchy case, the best-fit
value with a 68\%~CL is
$\textrm{sin}^{2}2\theta_{13} = 0.136^{+0.044}_{-0.033}$
($0.166^{+0.051}_{-0.042}$).
With the current statistics, the correlation between the $\nu_\mu$ disappearance and $\nu_{e}$ appearance measurements
%the correlation between the systematic
%errors of the $\nu_\mu$ disappearance and $\nu_{e}$ appearance
%measurements 
in T2K is negligibly small.
%The result is consistent with previous measurements.
%The value of sin$^2\theta_{23}$ is fixed to 0.4 (dashed line), 0.5 (solid
%line) and 0.6 (chain line). The value of sin$^2\theta_{23}$ largely
%affects the shape of the contours.

\begin{figure}[tbp]
 \includegraphics[width=0.45\textwidth]{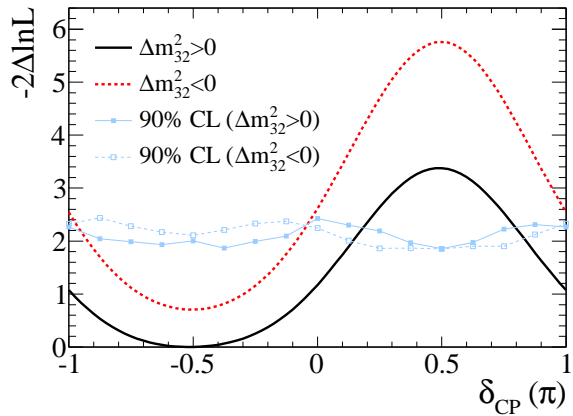}
  \caption{\label{fig:chi2_delta} 
  The $-2\Delta \ln{\cal L}$ value as a function of $\delta_{\mathrm{CP}}$ for
 normal hierarchy (solid line) and inverted hierarchy (dotted line). The likelihood is marginalized over sin$^22\theta_{13}$, sin$^2\theta_{23}$
 and $\Delta m^{2}_{32}$. The solid (dotted) line with markers 
 corresponds to the 90\% CL limits for normal (inverted) hierarchy, evaluated 
 by using the Feldman-Cousins method.
 The $\delta_{\mathrm{CP}}$ regions with values above the lines are excluded at 90\%~CL.
  }
\end{figure}

Constraints on $\delta_{\mathrm{CP}}$ are obtained by combining our results with the
$\theta_{13}$ value measured by reactor experiments. The additional likelihood
constraint term on sin$^22\theta_{13}$ is defined as
% TN 2013/10/28
%exp($-2($sin$^22\theta_{13}-0.098)^2/0.013^2$), 
$\exp\{-(\sin^2 2\theta_{13} -0.098)^2/(2(0.013^2))\}$,
where 0.098 and 0.013 are the averaged
value and the error of sin$^22\theta_{13}$ from PDG2012~\cite{Beringer:1900zz}.
The $-2\Delta \ln{\cal L}$ curve as a function of $\delta_{\mathrm{CP}}$ is shown in Figure~\ref{fig:chi2_delta}, where the likelihood is
%It is possible to extract constraints on $\delta_{\mathrm{CP}}$
%by combining these results with 
%reactor $\overline{\nu}_e$ disappearance measurements. 
%the average $\theta_{13}$ value from the PDG2012.
%Figure~\ref{fig:chi2_delta} shows a $-2\Delta \ln {\cal L}$ curve as a
%function of $\delta_{\mathrm{CP}}$ where the likelihood is 
marginalized over sin$^22\theta_{13}$, sin$^2\theta_{23}$ and $\Delta m^2_{32}$.
%The constraint is obtained using sin$^22\theta_{13}=0.098\pm0.013$ from PDG2012~\cite{Beringer:1900zz}
%and the T2K $\nu_\mu$ disappearance result~\cite{Abe:2013fuq}.
%The measurement constrains $\delta_{\mathrm{CP}}$ to be outside the region of 0.69 to 2.63~radians
%in the case of normal hierarchy and 0.27 and 3.06~radians for the
%inverted case at 90\% CL.
The combined T2K and reactor measurements prefer $\delta_{\mathrm{CP}} = -\pi/2$.
The 90\% CL limits shown in Figure \ref{fig:chi2_delta} are evaluated by using the Feldman-Cousins method~\cite{Feldman:1997qc}
in order to extract the excluded region. 
The data excludes $\delta_{\mathrm{CP}}$ between 0.19$\pi$ and 0.80$\pi$
($-\pi$ and $-0.97 \pi$, and $-0.04 \pi$ and $\pi$) with normal
(inverted) hierarchy at 90\% CL.
%This measurement excludes $\delta_{\mathrm{CP}}$ values between $0.19
%\pi$ and $0.80 \pi$ ($-\pi$ and $-0.97 \pi$, $-0.04 \pi$ and $\pi$) 
%assuming normal (inverted) hierarchy at 90\% CL.
%0.69 and 2.63~radians (0.27 and 3.06~radians)

The maximum value of $-2\Delta \ln{\cal L}$ is 3.38 (5.76) at $\delta_{\mathrm{CP}}=\pi/2$
for normal (inverted) hierarchy case.
This value is compared with a large number of toy MC
experiments, generated assuming $\delta_{\mathrm{CP}}=-\pi/2$, sin$^22\theta_{13}=0.1$, 
sin$^2\theta_{23}=0.5$ and $\Delta m^{2}_{32}= 2.4 \times 10^{-3}~\mathrm{eV}^{2}$.
The MC averaged value of $-2\Delta \ln{\cal L}$ at $\delta_{\mathrm{CP}}=\pi/2$ is 2.20 (4.10) for normal (inverted)
hierarchy case, and the probability of obtaining a value greater or equal to the observed value is
34.1\% (33.4\%). With the same MC settings, the expected 90\% CL exclusion region is evaluated to be between 
0.35$\pi$ and 0.63$\pi$ (0.09$\pi$ and 0.90$\pi$) radians for normal (inverted) hierarchy case.

{\it Conclusions}\textemdash
%In conclusion, 
T2K has made the first observation of electron neutrino appearance in a muon neutrino beam with a peak energy of 0.6~GeV and a baseline of 295~km. With the fixed parameters $|\Delta m^2_{32}| = 2.4 \times 10^{-3}$~$\rm eV^2$, $\sin^2 \theta_{23} = 0.5$, $\delta_{\mathrm{CP}}=0$, and  $\Delta m^2_{32} >0$ ($\Delta m^2_{32} <0$), a best-fit value of $\sin^2 2 \theta_{13}$ = $0.140^{+0.038}_{-0.032}$ ($0.170^{+0.045}_{-0.037}$) is obtained, with a significance of 7.3$\sigma$ over the hypothesis of $\sin^2 2 \theta_{13}=0$. 
When combining the T2K result with the world average value of $\theta_{13}$ from reactor experiments, 
some values of $\delta_{\mathrm{CP}}$ are disfavored at the 90\%~CL.

%The significance of 28 observed electron neutrino events 
%corresponds to $7.3 \sigma$ when compared to $4.64 \pm 0.53$ expected background events. 
%With $|\Delta m^2_{32}| = 2.4 \times 10^{-3}$~$\rm eV^2$, $\theta_{23} = \pi/4$, $\delta_{\mathrm{CP}}=0$, and 
%$\Delta m^2_{32} >0$ ($\Delta m^2_{32} <0$), 
%the best fit value of $\sin^2 2 \theta_{13}$ is $0.150^{+0.039}_{-0.034}$ ($0.182^{+0.046}_{-0.040}$).
%Marginalizing over $\theta_{23}$ and $\Delta m^2_{32}$, 
%$\sin^2 2 \theta_{13}$ is $0.146^{+0.045}_{-0.036}$ ($0.176^{+0.053}_{-0.043}$).
%Further marginalizing over $\theta_{13}$, 
%$\delta_{\mathrm{CP}}$ values between $0.22 \pi$ and $0.84 \pi$ ($0.09 \pi$ and $3.1 \pi$) radians are excluded
%assuming normal (inverted) hierarchy at 90\% CL.
%
T2K will continue to take data to measure the neutrino oscillation parameters
more precisely and to further explore CP violation in the lepton sector.

We thank the J-PARC staff for superb accelerator performance and the CERN NA61 collaboration for providing valuable particle production data.
We acknowledge the support of MEXT, Japan; 
NSERC, NRC and CFI, Canada;
CEA and CNRS/IN2P3, France;
DFG, Germany; 
INFN, Italy;
Ministry of Science and Higher Education, Poland;
RAS, RFBR and MES, Russia; 
MICINN and CPAN, Spain;
SNSF and SER, Switzerland;
STFC, U.K.; and 
DOE, U.S.A.
We also thank CERN for the UA1/NOMAD magnet, 
DESY for the HERA-B magnet mover system,
NII for SINET4,  the WestGrid and SciNet consortia in Compute Canada, and GridPP, UK.
In addition participation of individual researchers
and institutions has been further supported by funds from: ERC (FP7), EU; 
JSPS, Japan; 
Royal Society, UK; 
DOE Early Career program, U.S.A.

\bibliographystyle{apsrev4-1}
\bibliography{nue2013}% Produces the bibliography via BibTeX.

\providecommand{\noopsort}[1]{}\providecommand{\singleletter}[1]{#1}%
\begin{thebibliography}{31}%
\makeatletter
\providecommand \@ifxundefined [1]{%
 \@ifx{#1\undefined}
}%
\providecommand \@ifnum [1]{%
 \ifnum #1\expandafter \@firstoftwo
 \else \expandafter \@secondoftwo
 \fi
}%
\providecommand \@ifx [1]{%
 \ifx #1\expandafter \@firstoftwo
 \else \expandafter \@secondoftwo
 \fi
}%
\providecommand \natexlab [1]{#1}%
\providecommand \enquote  [1]{``#1''}%
\providecommand \bibnamefont  [1]{#1}%
\providecommand \bibfnamefont [1]{#1}%
\providecommand \citenamefont [1]{#1}%
\providecommand \href@noop [0]{\@secondoftwo}%
\providecommand \href [0]{\begingroup \@sanitize@url \@href}%
\providecommand \@href[1]{\@@startlink{#1}\@@href}%
\providecommand \@@href[1]{\endgroup#1\@@endlink}%
\providecommand \@sanitize@url [0]{\catcode `\\12\catcode `\$12\catcode
  `\&12\catcode `\#12\catcode `\^12\catcode `\_12\catcode `\%12\relax}%
\providecommand \@@startlink[1]{}%
\providecommand \@@endlink[0]{}%
\providecommand \url  [0]{\begingroup\@sanitize@url \@url }%
\providecommand \@url [1]{\endgroup\@href {#1}{\urlprefix }}%
\providecommand \urlprefix  [0]{URL }%
\providecommand \Eprint [0]{\href }%
\providecommand \doibase [0]{http://dx.doi.org/}%
\providecommand \selectlanguage [0]{\@gobble}%
\providecommand \bibinfo  [0]{\@secondoftwo}%
\providecommand \bibfield  [0]{\@secondoftwo}%
\providecommand \translation [1]{[#1]}%
\providecommand \BibitemOpen [0]{}%
\providecommand \bibitemStop [0]{}%
\providecommand \bibitemNoStop [0]{.\EOS\space}%
\providecommand \EOS [0]{\spacefactor3000\relax}%
\providecommand \BibitemShut  [1]{\csname bibitem#1\endcsname}%
\let\auto@bib@innerbib\@empty
%</preamble>
\bibitem [{\citenamefont {Fukuda}\ \emph {et~al.}(1998)\citenamefont {Fukuda}
  \emph {et~al.}}]{Fukuda:1998mi}%
  \BibitemOpen
  \bibfield  {author} {\bibinfo {author} {\bibfnamefont {Y.}~\bibnamefont
  {Fukuda}} \emph {et~al.} (\bibinfo {collaboration} {Super-Kamiokande
  Collaboration}),\ }\href {\doibase 10.1103/PhysRevLett.81.1562} {\bibfield
  {journal} {\bibinfo  {journal} {Phys.Rev.Lett.}\ }\textbf {\bibinfo {volume}
  {81}},\ \bibinfo {pages} {1562} (\bibinfo {year} {1998})}\BibitemShut
  {NoStop}%
\bibitem [{\citenamefont {Abe}\ \emph {et~al.}(2011{\natexlab{a}})\citenamefont
  {Abe} \emph {et~al.}}]{Abe:2011ks}%
  \BibitemOpen
  \bibfield  {author} {\bibinfo {author} {\bibfnamefont {K.}~\bibnamefont
  {Abe}} \emph {et~al.} (\bibinfo {collaboration} {T2K Collaboration}),\ }\href
  {\doibase 10.1016/j.nima.2011.06.067} {\bibfield  {journal} {\bibinfo
  {journal} {Nucl.Instrum.Meth.}\ }\textbf {\bibinfo {volume} {A659}},\
  \bibinfo {pages} {106} (\bibinfo {year} {2011}{\natexlab{a}})}\BibitemShut
  {NoStop}%
%%CITATION = ARXIV:1106.1238;%%
\bibitem [{\citenamefont {Abe}\ \emph {et~al.}(2011{\natexlab{b}})\citenamefont
  {Abe} \emph {et~al.}}]{Abe:2011sj}%
  \BibitemOpen
  \bibfield  {author} {\bibinfo {author} {\bibfnamefont {K.}~\bibnamefont
  {Abe}} \emph {et~al.} (\bibinfo {collaboration} {T2K Collaboration}),\ }\href
  {\doibase 10.1103/PhysRevLett.107.041801} {\bibfield  {journal} {\bibinfo
  {journal} {Phys.Rev.Lett.}\ }\textbf {\bibinfo {volume} {107}},\ \bibinfo
  {pages} {041801} (\bibinfo {year} {2011}{\natexlab{b}})}\BibitemShut
  {NoStop}%
%%CITATION = ARXIV:1106.2822;%%
\bibitem [{\citenamefont {Abe}\ \emph {et~al.}(2013{\natexlab{a}})\citenamefont
  {Abe} \emph {et~al.}}]{Abe:2013xua}%
  \BibitemOpen
  \bibfield  {author} {\bibinfo {author} {\bibfnamefont {K.}~\bibnamefont
  {Abe}} \emph {et~al.} (\bibinfo {collaboration} {T2K Collaboration}),\ }\href
  {\doibase 10.1103/PhysRevD.88.032002} {\bibfield  {journal} {\bibinfo
  {journal} {Phys.Rev.}\ }\textbf {\bibinfo {volume} {D88}},\ \bibinfo {pages}
  {032002} (\bibinfo {year} {2013}{\natexlab{a}})}\BibitemShut {NoStop}%
\bibitem [{\citenamefont {Maki}\ \emph {et~al.}(1962)\citenamefont {Maki},
  \citenamefont {Nakagawa},\ and\ \citenamefont {Sakata}}]{Maki:1962mu}%
  \BibitemOpen
  \bibfield  {author} {\bibinfo {author} {\bibfnamefont {Z.}~\bibnamefont
  {Maki}}, \bibinfo {author} {\bibfnamefont {M.}~\bibnamefont {Nakagawa}}, \
  and\ \bibinfo {author} {\bibfnamefont {S.}~\bibnamefont {Sakata}},\ }\href
  {\doibase 10.1143/PTP.28.870} {\bibfield  {journal} {\bibinfo  {journal}
  {Prog.Theor.Phys.}\ }\textbf {\bibinfo {volume} {28}},\ \bibinfo {pages}
  {870} (\bibinfo {year} {1962})}\BibitemShut {NoStop}%
%%CITATION = PTPKA,28,870;%%
\bibitem [{\citenamefont {Pontecorvo}(1968)}]{Pontecorvo:1967fh}%
  \BibitemOpen
  \bibfield  {author} {\bibinfo {author} {\bibfnamefont {B.}~\bibnamefont
  {Pontecorvo}},\ }\href@noop {} {\bibfield  {journal} {\bibinfo  {journal}
  {Sov.Phys.JETP}\ }\textbf {\bibinfo {volume} {26}},\ \bibinfo {pages} {984}
  (\bibinfo {year} {1968})}\BibitemShut {NoStop}%
%%CITATION = SPHJA,26,984;%%
\bibitem [{\citenamefont {Freund}(2001)}]{Freund:2001pn}%
  \BibitemOpen
  \bibfield  {author} {\bibinfo {author} {\bibfnamefont {M.}~\bibnamefont
  {Freund}},\ }\href {\doibase 10.1103/PhysRevD.64.053003} {\bibfield
  {journal} {\bibinfo  {journal} {Phys.Rev.}\ }\textbf {\bibinfo {volume}
  {D64}},\ \bibinfo {pages} {053003} (\bibinfo {year} {2001})}\BibitemShut
  {NoStop}%
\bibitem [{\citenamefont {Beringer}\ \emph {et~al.}(2012)\citenamefont
  {Beringer} \emph {et~al.}}]{Beringer:1900zz}%
  \BibitemOpen
  \bibfield  {author} {\bibinfo {author} {\bibfnamefont {J.}~\bibnamefont
  {Beringer}} \emph {et~al.} (\bibinfo {collaboration} {Particle Data Group}),\
  }\href {\doibase 10.1103/PhysRevD.86.010001} {\bibfield  {journal} {\bibinfo
  {journal} {Phys.Rev.}\ }\textbf {\bibinfo {volume} {D86}},\ \bibinfo {pages}
  {010001} (\bibinfo {year} {2012})}\BibitemShut {NoStop}%
%%CITATION = PHRVA,D86,010001;%%
\bibitem [{\citenamefont {An}\ \emph {et~al.}(2012)\citenamefont {An} \emph
  {et~al.}}]{An:2012eh}%
  \BibitemOpen
  \bibfield  {author} {\bibinfo {author} {\bibfnamefont {F.}~\bibnamefont {An}}
  \emph {et~al.} (\bibinfo {collaboration} {Daya Bay Collaboration}),\ }\href
  {\doibase 10.1103/PhysRevLett.108.171803} {\bibfield  {journal} {\bibinfo
  {journal} {Phys.Rev.Lett.}\ }\textbf {\bibinfo {volume} {108}},\ \bibinfo
  {pages} {171803} (\bibinfo {year} {2012})}\BibitemShut {NoStop}%
%%CITATION = ARXIV:1203.1669;%%
\bibitem [{\citenamefont {Ahn}\ \emph {et~al.}(2012)\citenamefont {Ahn} \emph
  {et~al.}}]{Ahn:2012nd}%
  \BibitemOpen
  \bibfield  {author} {\bibinfo {author} {\bibfnamefont {J.}~\bibnamefont
  {Ahn}} \emph {et~al.} (\bibinfo {collaboration} {RENO collaboration}),\
  }\href {\doibase 10.1103/PhysRevLett.108.191802} {\bibfield  {journal}
  {\bibinfo  {journal} {Phys.Rev.Lett.}\ }\textbf {\bibinfo {volume} {108}},\
  \bibinfo {pages} {191802} (\bibinfo {year} {2012})}\BibitemShut {NoStop}%
%%CITATION = ARXIV:1204.0626;%%
\bibitem [{\citenamefont {Abe}\ \emph {et~al.}(2012{\natexlab{a}})\citenamefont
  {Abe} \emph {et~al.}}]{Abe:2011fz}%
  \BibitemOpen
  \bibfield  {author} {\bibinfo {author} {\bibfnamefont {Y.}~\bibnamefont
  {Abe}} \emph {et~al.} (\bibinfo {collaboration} {Double Chooz
  Collaboration}),\ }\href {\doibase 10.1103/PhysRevLett.108.131801} {\bibfield
   {journal} {\bibinfo  {journal} {Phys.Rev.Lett.}\ }\textbf {\bibinfo {volume}
  {108}},\ \bibinfo {pages} {131801} (\bibinfo {year}
  {2012}{\natexlab{a}})}\BibitemShut {NoStop}%
%%CITATION = ARXIV:1112.6353;%%
\bibitem [{\citenamefont {Adamson}\ \emph {et~al.}(2013)\citenamefont {Adamson}
  \emph {et~al.}}]{Adamson:2013ue}%
  \BibitemOpen
  \bibfield  {author} {\bibinfo {author} {\bibfnamefont {P.}~\bibnamefont
  {Adamson}} \emph {et~al.} (\bibinfo {collaboration} {MINOS Collaboration}),\
  }\href {\doibase 10.1103/PhysRevLett.110.171801} {\bibfield  {journal}
  {\bibinfo  {journal} {Phys.Rev.Lett.}\ }\textbf {\bibinfo {volume} {110}},\
  \bibinfo {pages} {171801} (\bibinfo {year} {2013})}\BibitemShut {NoStop}%
\bibitem [{\citenamefont {Abe}\ \emph {et~al.}(2013{\natexlab{b}})\citenamefont
  {Abe} \emph {et~al.}}]{Abe:2012av}%
  \BibitemOpen
  \bibfield  {author} {\bibinfo {author} {\bibfnamefont {K.}~\bibnamefont
  {Abe}} \emph {et~al.} (\bibinfo {collaboration} {T2K Collaboration}),\ }\href
  {\doibase 10.1103/PhysRevD.87.012001, 10.1103/PhysRevD.87.019902} {\bibfield
  {journal} {\bibinfo  {journal} {Phys.Rev.}\ }\textbf {\bibinfo {volume}
  {D87}},\ \bibinfo {pages} {012001} (\bibinfo {year}
  {2013}{\natexlab{b}})}\BibitemShut {NoStop}%
%%CITATION = ARXIV:1211.0469;%%
\bibitem [{\citenamefont {Abgrall}\ \emph
  {et~al.}(2011{\natexlab{a}})\citenamefont {Abgrall} \emph
  {et~al.}}]{Abgrall:2011ae}%
  \BibitemOpen
  \bibfield  {author} {\bibinfo {author} {\bibfnamefont {N.}~\bibnamefont
  {Abgrall}} \emph {et~al.} (\bibinfo {collaboration} {NA61/SHINE
  Collaboration}),\ }\href {\doibase 10.1103/PhysRevC.84.034604} {\bibfield
  {journal} {\bibinfo  {journal} {Phys.Rev.}\ }\textbf {\bibinfo {volume}
  {C84}},\ \bibinfo {pages} {034604} (\bibinfo {year}
  {2011}{\natexlab{a}})}\BibitemShut {NoStop}%
%%CITATION = ARXIV:1102.0983;%%
\bibitem [{\citenamefont {Abgrall}\ \emph {et~al.}(2012)\citenamefont {Abgrall}
  \emph {et~al.}}]{Abgrall:2011ts}%
  \BibitemOpen
  \bibfield  {author} {\bibinfo {author} {\bibfnamefont {N.}~\bibnamefont
  {Abgrall}} \emph {et~al.} (\bibinfo {collaboration} {NA61/SHINE
  Collaboration}),\ }\href {\doibase 10.1103/PhysRevC.85.035210} {\bibfield
  {journal} {\bibinfo  {journal} {Phys.Rev.}\ }\textbf {\bibinfo {volume}
  {C85}},\ \bibinfo {pages} {035210} (\bibinfo {year} {2012})}\BibitemShut
  {NoStop}%
%%CITATION = ARXIV:1112.0150;%%
\bibitem [{\citenamefont {Ferrari}\ \emph {et~al.}()\citenamefont {Ferrari},
  \citenamefont {Sala}, \citenamefont {Fasso},\ and\ \citenamefont
  {Ranft}}]{FLUKA}%
  \BibitemOpen
  \bibfield  {author} {\bibinfo {author} {\bibfnamefont {A.}~\bibnamefont
  {Ferrari}}, \bibinfo {author} {\bibfnamefont {P.~R.}\ \bibnamefont {Sala}},
  \bibinfo {author} {\bibfnamefont {A.}~\bibnamefont {Fasso}}, \ and\ \bibinfo
  {author} {\bibfnamefont {J.}~\bibnamefont {Ranft}},\ }\href@noop {} {\bibinfo
   {journal} {CERN-2005-010, SLAC-R-773, INFN-TC-05-11}\ }\BibitemShut
  {NoStop}%
\bibitem [{\citenamefont {Battistoni}\ \emph {et~al.}(2007)\citenamefont
  {Battistoni}, \citenamefont {Muraro}, \citenamefont {Sala}, \citenamefont
  {Cerutti}, \citenamefont {Ferrari} \emph {et~al.}}]{FLUKA08}%
  \BibitemOpen
\bibfield  {journal} {  }\bibfield  {author} {\bibinfo {author} {\bibfnamefont
  {G.}~\bibnamefont {Battistoni}}, \bibinfo {author} {\bibfnamefont
  {S.}~\bibnamefont {Muraro}}, \bibinfo {author} {\bibfnamefont {P.~R.}\
  \bibnamefont {Sala}}, \bibinfo {author} {\bibfnamefont {F.}~\bibnamefont
  {Cerutti}}, \bibinfo {author} {\bibfnamefont {A.}~\bibnamefont {Ferrari}},
  \emph {et~al.},\ }\href {\doibase 10.1063/1.2720455} {\bibfield  {journal}
  {\bibinfo  {journal} {AIP Conf.Proc.}\ }\textbf {\bibinfo {volume} {896}},\
  \bibinfo {pages} {31} (\bibinfo {year} {2007})}\BibitemShut {NoStop}%
%%CITATION = APCPC,896,31;%%
\bibitem [{\citenamefont {Brun}\ \emph {et~al.}(1994)\citenamefont {Brun},
  \citenamefont {Carminati},\ and\ \citenamefont {Giani}}]{GEANT3}%
  \BibitemOpen
  \bibfield  {author} {\bibinfo {author} {\bibfnamefont {R.}~\bibnamefont
  {Brun}}, \bibinfo {author} {\bibfnamefont {F.}~\bibnamefont {Carminati}}, \
  and\ \bibinfo {author} {\bibfnamefont {S.}~\bibnamefont {Giani}},\
  }\href@noop {} {\bibfield  {journal} {\bibinfo  {journal} {CERN-W5013}\ }
  (\bibinfo {year} {1994})}\BibitemShut {NoStop}%
\bibitem [{\citenamefont {Zeitnitz}\ and\ \citenamefont
  {Gabriel}(1992)}]{Zeitnitz:1992vw}%
  \BibitemOpen
  \bibfield  {author} {\bibinfo {author} {\bibfnamefont {C.}~\bibnamefont
  {Zeitnitz}}\ and\ \bibinfo {author} {\bibfnamefont {T.}~\bibnamefont
  {Gabriel}},\ }\href@noop {} {\bibfield  {journal} {\bibinfo  {journal} {In
  *Corpus Christi 1992, Calorimetry in high energy physics* 394-404}\ }
  (\bibinfo {year} {1992})}\BibitemShut {NoStop}%
%%CITATION = INSPIRE-348996;%%
\bibitem [{\citenamefont {Abe}\ \emph {et~al.}(2012{\natexlab{b}})\citenamefont
  {Abe} \emph {et~al.}}]{Abe:2011xv}%
  \BibitemOpen
  \bibfield  {author} {\bibinfo {author} {\bibfnamefont {K.}~\bibnamefont
  {Abe}} \emph {et~al.} (\bibinfo {collaboration} {T2K Collaboration}),\ }\href
  {\doibase 10.1016/j.nima.2012.03.023} {\bibfield  {journal} {\bibinfo
  {journal} {Nucl.Instrum.Meth.}\ }\textbf {\bibinfo {volume} {A694}},\
  \bibinfo {pages} {211} (\bibinfo {year} {2012}{\natexlab{b}})}\BibitemShut
  {NoStop}%
\bibitem [{\citenamefont {Hayato}(2009)}]{Hayato:2009}%
  \BibitemOpen
  \bibfield  {author} {\bibinfo {author} {\bibfnamefont {Y.}~\bibnamefont
  {Hayato}},\ }\href@noop {} {\bibfield  {journal} {\bibinfo  {journal} {Acta
  Phys.Polon.}\ }\textbf {\bibinfo {volume} {B40}},\ \bibinfo {pages} {2477}
  (\bibinfo {year} {2009})}\BibitemShut {NoStop}%
%%CITATION = APPOA,B40,2477;%%
\bibitem [{\citenamefont {Aguilar-Arevalo}\ \emph
  {et~al.}(2011{\natexlab{a}})\citenamefont {Aguilar-Arevalo} \emph
  {et~al.}}]{mb-cc1pi0}%
  \BibitemOpen
  \bibfield  {author} {\bibinfo {author} {\bibfnamefont {A.~A.}\ \bibnamefont
  {Aguilar-Arevalo}} \emph {et~al.} (\bibinfo {collaboration} {MiniBooNE
  Collaboration}),\ }\href {\doibase 10.1103/PhysRevD.83.052009} {\bibfield
  {journal} {\bibinfo  {journal} {Phys.Rev.}\ }\textbf {\bibinfo {volume}
  {D83}},\ \bibinfo {pages} {052009} (\bibinfo {year}
  {2011}{\natexlab{a}})}\BibitemShut {NoStop}%
\bibitem [{\citenamefont {Aguilar-Arevalo}\ \emph
  {et~al.}(2011{\natexlab{b}})\citenamefont {Aguilar-Arevalo} \emph
  {et~al.}}]{mb-cc1pip}%
  \BibitemOpen
  \bibfield  {author} {\bibinfo {author} {\bibfnamefont {A.~A.}\ \bibnamefont
  {Aguilar-Arevalo}} \emph {et~al.} (\bibinfo {collaboration} {MiniBooNE
  Collaboration}),\ }\href {\doibase 10.1103/PhysRevD.83.052007} {\bibfield
  {journal} {\bibinfo  {journal} {Phys.Rev.}\ }\textbf {\bibinfo {volume}
  {D83}},\ \bibinfo {pages} {052007} (\bibinfo {year}
  {2011}{\natexlab{b}})}\BibitemShut {NoStop}%
\bibitem [{\citenamefont {Aguilar-Arevalo}\ \emph {et~al.}(2010)\citenamefont
  {Aguilar-Arevalo} \emph {et~al.}}]{mb-nc1pi0}%
  \BibitemOpen
  \bibfield  {author} {\bibinfo {author} {\bibfnamefont {A.~A.}\ \bibnamefont
  {Aguilar-Arevalo}} \emph {et~al.} (\bibinfo {collaboration} {MiniBooNE
  Collaboration}),\ }\href {\doibase 10.1103/PhysRevD.81.013005} {\bibfield
  {journal} {\bibinfo  {journal} {Phys.Rev.}\ }\textbf {\bibinfo {volume}
  {D81}},\ \bibinfo {pages} {013005} (\bibinfo {year} {2010})}\BibitemShut
  {NoStop}%
%%CITATION = ARXIV:0911.2063;%%
\bibitem [{\citenamefont {Amaudruz}\ \emph {et~al.}(2012)\citenamefont
  {Amaudruz} \emph {et~al.}}]{Amaudruz:2012pe}%
  \BibitemOpen
  \bibfield  {author} {\bibinfo {author} {\bibfnamefont {P.}~\bibnamefont
  {Amaudruz}} \emph {et~al.} (\bibinfo {collaboration} {T2K ND280 FGD group}),\
  }\href {\doibase 10.1016/j.nima.2012.08.020} {\bibfield  {journal} {\bibinfo
  {journal} {Nucl.Instrum.Meth.}\ }\textbf {\bibinfo {volume} {A696}},\
  \bibinfo {pages} {1} (\bibinfo {year} {2012})}\BibitemShut {NoStop}%
%%CITATION = ARXIV:1204.3666;%%
\bibitem [{\citenamefont {Abgrall}\ \emph
  {et~al.}(2011{\natexlab{b}})\citenamefont {Abgrall} \emph
  {et~al.}}]{Abgrall:2010hi}%
  \BibitemOpen
  \bibfield  {author} {\bibinfo {author} {\bibfnamefont {N.}~\bibnamefont
  {Abgrall}} \emph {et~al.} (\bibinfo {collaboration} {T2K ND280 TPC group}),\
  }\href {\doibase 10.1016/j.nima.2011.02.036} {\bibfield  {journal} {\bibinfo
  {journal} {Nucl.Instrum.Meth.}\ }\textbf {\bibinfo {volume} {A637}},\
  \bibinfo {pages} {25} (\bibinfo {year} {2011}{\natexlab{b}})}\BibitemShut
  {NoStop}%
%%CITATION = ARXIV:1012.0865;%%
\bibitem [{\citenamefont {Patterson}\ \emph {et~al.}(2009)\citenamefont
  {Patterson} \emph {et~al.}}]{miniboonerecon}%
  \BibitemOpen
  \bibfield  {author} {\bibinfo {author} {\bibfnamefont {R.~B.}\ \bibnamefont
  {Patterson}} \emph {et~al.},\ }\href@noop {} {\bibfield  {journal} {\bibinfo
  {journal} {Nucl.Instrum.Meth.}\ }\textbf {\bibinfo {volume} {A608}},\
  \bibinfo {pages} {206} (\bibinfo {year} {2009})}\BibitemShut {NoStop}%
\bibitem [{\citenamefont {Abe}\ \emph {et~al.}(2013{\natexlab{c}})\citenamefont
  {Abe} \emph {et~al.}}]{t2knue}%
  \BibitemOpen
  \bibfield  {author} {\bibinfo {author} {\bibfnamefont {K.}~\bibnamefont
  {Abe}} \emph {et~al.} (\bibinfo {collaboration} {T2K Collaboration}),\
  }\href@noop {} {\bibfield  {journal} {\bibinfo  {journal} {Phys.Rev.}\
  }\textbf {\bibinfo {volume} {D88}},\ \bibinfo {pages} {032002} (\bibinfo
  {year} {2013}{\natexlab{c}})}\BibitemShut {NoStop}%
\bibitem [{\citenamefont {Fogli}\ \emph {et~al.}(2011)\citenamefont {Fogli},
  \citenamefont {Lisi}, \citenamefont {Marrone}, \citenamefont {Palazzo},\ and\
  \citenamefont {Rotunno}}]{Fogli}%
  \BibitemOpen
  \bibfield  {author} {\bibinfo {author} {\bibfnamefont {G.~L.}\ \bibnamefont
  {Fogli}}, \bibinfo {author} {\bibfnamefont {E.}~\bibnamefont {Lisi}},
  \bibinfo {author} {\bibfnamefont {A.}~\bibnamefont {Marrone}}, \bibinfo
  {author} {\bibfnamefont {A.}~\bibnamefont {Palazzo}}, \ and\ \bibinfo
  {author} {\bibfnamefont {A.~M.}\ \bibnamefont {Rotunno}},\ }\href {\doibase
  10.1103/PhysRevD.84.053007} {\bibfield  {journal} {\bibinfo  {journal}
  {Phys.Rev.}\ }\textbf {\bibinfo {volume} {D84}},\ \bibinfo {pages} {053007}
  (\bibinfo {year} {2011})}\BibitemShut {NoStop}%
\bibitem [{\citenamefont {Abe}\ \emph {et~al.}(2013{\natexlab{d}})\citenamefont
  {Abe} \emph {et~al.}}]{Abe:2013fuq}%
  \BibitemOpen
  \bibfield  {author} {\bibinfo {author} {\bibfnamefont {K.}~\bibnamefont
  {Abe}} \emph {et~al.} (\bibinfo {collaboration} {T2K Collaboration}),\
  }\href@noop {} {\  (\bibinfo {year} {2013}{\natexlab{d}})},\ \Eprint
  {http://arxiv.org/abs/1308.0465} {arXiv:1308.0465 [hep-ex]} \BibitemShut
  {NoStop}%
%%CITATION = ARXIV:1308.0465;%%
\bibitem [{\citenamefont {Feldman}\ and\ \citenamefont
  {Cousins}(1998)}]{Feldman:1997qc}%
  \BibitemOpen
  \bibfield  {author} {\bibinfo {author} {\bibfnamefont {G.~J.}\ \bibnamefont
  {Feldman}}\ and\ \bibinfo {author} {\bibfnamefont {R.~D.}\ \bibnamefont
  {Cousins}},\ }\href {\doibase 10.1103/PhysRevD.57.3873} {\bibfield  {journal}
  {\bibinfo  {journal} {Phys.Rev.}\ }\textbf {\bibinfo {volume} {D57}},\
  \bibinfo {pages} {3873} (\bibinfo {year} {1998})}\BibitemShut {NoStop}%
\end{thebibliography}%

\end{document}